\documentclass[aip,pof,reprint,groupedaddress]{revtex4-1}

\usepackage{psfrag,natbib}
\usepackage{graphicx}
\usepackage{amsmath}
\usepackage{amssymb}
\usepackage{wasysym}
\usepackage{color}
\usepackage{xcolor}
\usepackage{tikz}
\usepackage{subfigure}

\usepackage{docs}%
\usepackage{bm}%
\usepackage[colorlinks=true,linkcolor=blue]{hyperref}%
\expandafter\ifx\csname package@font\endcsname\relax\else
 \expandafter\expandafter
 \expandafter\usepackage
 \expandafter\expandafter
 \expandafter{\csname package@font\endcsname}%
\fi
\newcommand{\mean}[1]{\left\langle #1 \right\rangle}

\newcommand{\dd}{{\,\rm d}}

\begin{document}

\title{The temporal evolution of the energy flux across scales 
in homogeneous turbulence}%

\author{J.I. Cardesa}
\email[]{jose@torroja.dmt.upm.es}
\author{A. Vela-Mart\'in}
\author{S. Dong}
\author{J. Jim\'enez}
\affiliation{School of Aeronautics, Universidad Polit\'ecnica de Madrid, 28040 Madrid, Spain}

\date{\today}
\begin{abstract}
A temporal study of energy transfer across length scales is performed in 3D numerical
simulations of homogeneous shear flow and isotropic turbulence. The average time taken 
by perturbations in the energy flux to travel between scales is measured and shown to 
be additive. Our data suggests that the propagation of disturbances in the
energy flux is independent of the forcing and that it defines a `velocity' that
determines the energy flux itself. These results support that the cascade is, on
average, a scale-local process where energy is continuously transmitted from one
scale to the next in order of decreasing size.
\end{abstract}


\maketitle

The difficulty in understanding a multiscale problem such as turbulence has been 
frequently tackled by assuming \textit{a priori} some type of simplified phenomenology. 
Take, for instance, Richardson's cartoon based on concepts such as eddies whose 
energy cascades by the successive breakup of larger eddies into smaller ones, until 
it is dissipated by viscosity.\cite{richardson2007weather} It was later 
used as the basis for the more quantitative work of Kolmogorov.\cite{K41} 
We now know that in 3D turbulence the energy does cascade towards the 
smallest scales,\cite{kraichnan1971cascade} at least on average. But several 
models have been discussed in the literature which are consistent with this average trend 
yet differ in their detailed mechanism. 
For example, the energy could jump directly from large eddies to much smaller ones, 
\cite{Lumley_somethoughts_PoF1992,Tsinober_tb} contradicting the scale locality 
assumed by Richardson, or include frequent excursions from smaller 
to larger scales -- a process coined backscatter  
-- questioning a unique directionality.\cite{piomelli_1991,Kaneda_ann_rev_2009}
Such alternative roads leading to the same 
average behavior urge an improved understanding of the cascade dynamics. In this
letter, we report on the time taken by \textcolor{black}{disturbances in the the energy flux to travel in scale space} -- 
an essential ingredient in unsteady phenomenological models.\cite{Lumley_somethoughts_PoF1992}
%
%
To overcome the difficulty in generating a wide 
dynamic range in space but especially in time, our direct numerical simulations are 
purposely run for very long times to perform a temporal cross-correlation 
analysis between energy fluxes at various length scales. 
\par Our approach is based on the large- and small-scale decomposition of the 
instantaneous velocity field according to 
$u_{i}(x_{i},t) = \widetilde{u}_{i}(x_{i},t) + u'_{i}(x_{i},t)$,\cite{obukhoff1941energy} 
where $\widetilde{u}_{i}$ is the \textit{i-th} component of the spatially low-pass 
filtered velocity. In an incompressible flow with kinematic viscosity $\nu$, 
the kinetic energy of the large-scale field evolves as
\begin{equation}
  \left( \frac{\partial}{\partial t}+ \widetilde{u}_{j}\frac{\partial}{\partial x_{j}}\right)  \frac{1}{2}\widetilde{u}_{i}\widetilde{u}_{i} = 
-\frac{\partial}{\partial x_{j}} \left(\widetilde{u}_{j}\widetilde{p} + 
\widetilde{u}_{i}\tau_{ij}-2\nu \widetilde{u}_{i} \widetilde{S}_{ij} \right)
  - 2\nu \widetilde{S}_{ij}\widetilde{S}_{ij}
  -\Sigma + \widetilde{u}_{i}\widetilde{f}_{i},   \label{eqn:kin_e_filt}
\end{equation}
where $\widetilde{S}_{ij}=\left(\partial \widetilde{u}_{i}/\partial x_{j} + \partial
\widetilde{u}_{j}/\partial x_{i}\right)/2$ is the strain-rate tensor of the
large-scale velocities, $\Sigma=-\tau_{ij}\widetilde{S}_{ij}$,
$\tau_{ij} = \widetilde{u_{i}u_{j}} -
\widetilde{u}_{i}\widetilde{u}_{j}$ is the subgrid-scale stress tensor and $f_{i}$
is the forcing term.
Eq.~(\ref{eqn:kin_e_filt}) has been studied extensively in the context of large-eddy
simulations (LES), where capturing the behavior of
$\Sigma$ is at the cornerstone of most modeling
strategies. \cite{Germano_Piomelli} In the mean, $\Sigma$ is positive and acts as a
net energy removal from the large scales by the small 
ones.\cite{piomelli_1991,Meneveau_cerutti_PoF1998} 
We choose it as our real-space
marker of cross-scale energy transfer, and study it in two
different flows: homogeneous shear turbulence (HST) and homogeneous isotropic
turbulence (HIT). The details of the simulations are given in
Table~\ref{tab:DNS_settings}. The low-pass filtered velocity in HIT was obtained 
by multiplying the Fourier modes $\hat{u}_{i}(\boldsymbol{k},t)$ by an isotropic 
Gaussian kernel 
$\hat{G}(\boldsymbol{k}) = \exp[-(r\boldsymbol{k})^{2}/24]$,
\cite{aoyama_2005,Pope_book} where $\boldsymbol{k}$ is 
the wavevector and $r$ the filter width. In HST this method could only 
be applied in the streamwise and spanwise directions, along which the 
flow was Fourier-discretized. Since seventh-order compact finite differences 
were used in the vertical (sheared) direction, filtering along it was 
implemented by convolving the velocity with the real-space
transform of $\hat{G}\left(\boldsymbol{k}\right)$:
$G\left(x\right) = P\exp[-6x^{2}/r^{2}]$, where $P$ is a constant chosen to 
meet the normalization condition. For comparison, a sharp spectral filter was 
used on some HIT fields with $\hat{G}(\boldsymbol{k}) = 0$ when 
$|\boldsymbol{k}|\geq \pi/r$, $\hat{G}(\boldsymbol{k}) = 1$ otherwise.
\par Other energy transfer markers exist which could have been suited 
to the HIT simulation. An obvious candidate comes from the spectral energy 
equation in isotropic turbulence 
\begin{equation}
  \frac{\partial E(k,t)}{\partial t} = F(k,t) - 2\nu k^{2}E(k,t) + \Xi(k,t),
\end{equation}
where $E\left(k,t \right)$ is the 3D instantaneous energy spectrum,
$k = |\boldsymbol{k}|$ \textcolor{black}{and $\Xi$ is the forcing.\cite{Hinze}} 
The spectral energy flux $\Pi(k)=\int_{0}^{k} F(k) \dd k$ 
is often invoked in energy cascade studies. For a single flow field of isotropic 
turbulence, $\Pi(k)$ and $\mean{\Sigma(r)}$ are equal when a sharp spectral filter
is used to compute $\Sigma$ with a cut-off wavenumber $k=\pi/r$.  
(Hereafter, $\mean{\theta}$ is the time-dependent spatial average of $\theta$ over the 
computational domain, while $\overline{\theta}$ is the temporal mean of $\mean{\theta}$.)
The difference between the two energy transfer markers thus amounts to a choice of
filter type. Since $\Pi$ is of limited use in flows other than HIT and given the 
relevance of $\Sigma$ in LES, we favored $\Sigma$ for comparison 
between the two flows.
\begin{table*}
\caption{Parameters of the simulations. $Re_{\lambda}$ is the Reynolds number 
based on the Taylor-microscale. $N_{i}$ and $L_{i}$ are the number of real
Fourier modes and the domain size in directions $i={x,y,z}$. Length scales are $\eta=\left(
\nu^{3}/\overline{\varepsilon}\right)^{1/4}$ and
$L_{o}=\overline{K}^{3/2}/\overline{\varepsilon}$. Times are normalized by
$T_{o}=\overline{K}/\overline{\varepsilon}$. $T_{simu}$ is the simulation time, and
$\Delta t_{tot}$ is the average delay between $\mean{K}$ and $\mean{\varepsilon}$.
$T_{KK}$ is an autocorrelation time for $\mean{K}$, defined in the text. $K=u_{i}u_{i}/2$.
\label{tab:DNS_settings}
}
  \begin{ruledtabular}
    \begin{tabular}{lccccccc}
      Case& $Re_{\lambda}$ & $N_{x}\hspace{-0.08cm}\times\hspace{-0.08cm}N_{y}
\hspace{-0.08cm}\times\hspace{-0.08cm}N_{z}$ & $(L_{x}\times L_{y} \times L_{z})/\eta$ & 
$L_{o}/\eta$
      & $T_{simu}/T_{o}$ & $\Delta t_{tot}/T_{o}$ & $T_{KK}/\Delta t_{tot}$ \\ \hline
      HST & 107 & $768\times512\times255$ & $1117\times745\times372$ & 267 & 213 & 0.49 
& 9.04\\
      HIT1 & 146 & $256^3$ & $ 506^3$ & 425 & 165 & 0.53 & 2.32\\
      HIT2 & 236 & $512^3$ & $ 1011^3$ & 876 & 16 & 0.40 & 2.36 \\
      HIT3 & 384 & $1024^3$ & $ 2022^3$ & 1813 & 4 & 0.42 & 2.14 \\ 
    \end{tabular}
  \end{ruledtabular}
\end{table*}
\begin{table*}[b!]
\caption{Ratio $\Phi_{F}/\Phi_{B}$ of forward to reverse energy flux. Numbers in
parentheses are the corresponding volume ratios, 
$\int_0^\infty \rho\left( \Sigma \right)\dd\Sigma/ \int_{-\infty}^{0} \rho\left( \Sigma \right)\dd\Sigma$ .%
\label{tab:predom_Phi_D}}
  \begin{ruledtabular}
    \begin{tabular}{lcccccc}
      $r/\eta$ & 8 &16 & 31 &  62 & 125 &  250 \\ 
      \hline
      HST(Gauss) & $N/A$ & 16(4) & 24(5) & 62(10) & 139(19) & $N/A$ \\
      HIT2(Gauss) & 15(3) & 15(4) & 21(5) & 39(8) & 59(10) & 37(9) \\
      HIT2(sharp) & 1.4(1.2) & 1.7(1.3) & 2.2(1.6) & 2.9(1.9) & 3.7(2.3) & 5.0(2.7) \\
    \end{tabular}
  \end{ruledtabular}
\end{table*}
\par
\begin{figure}[t!]
  \centering
  \begin{minipage}{\textwidth}
    \includegraphics[trim = 0cm 0cm 0cm
    0cm,width=5.2cm]{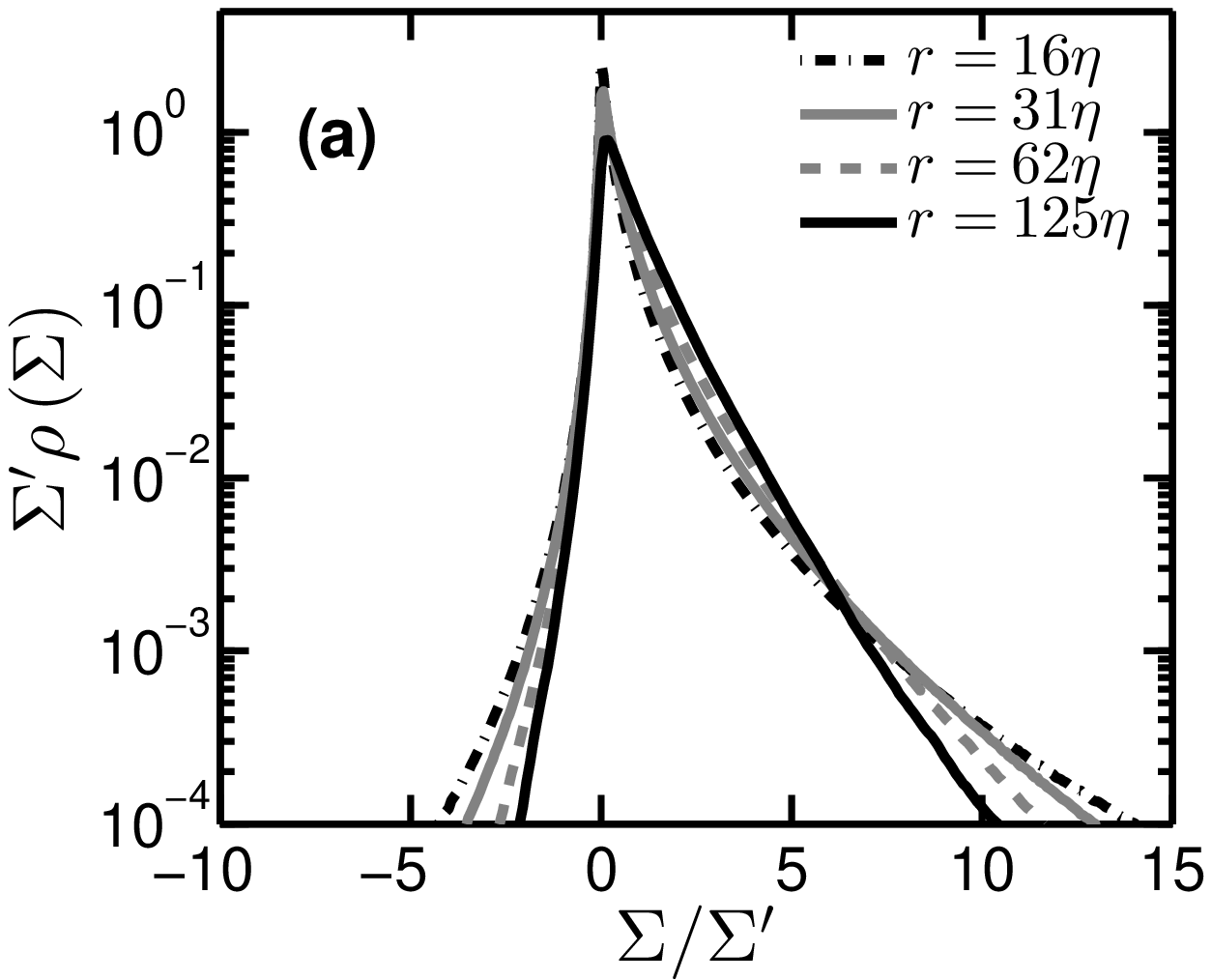}\hspace{1.5cm}\includegraphics[trim = 0cm 0cm 0cm
    0cm,width=5.2cm]{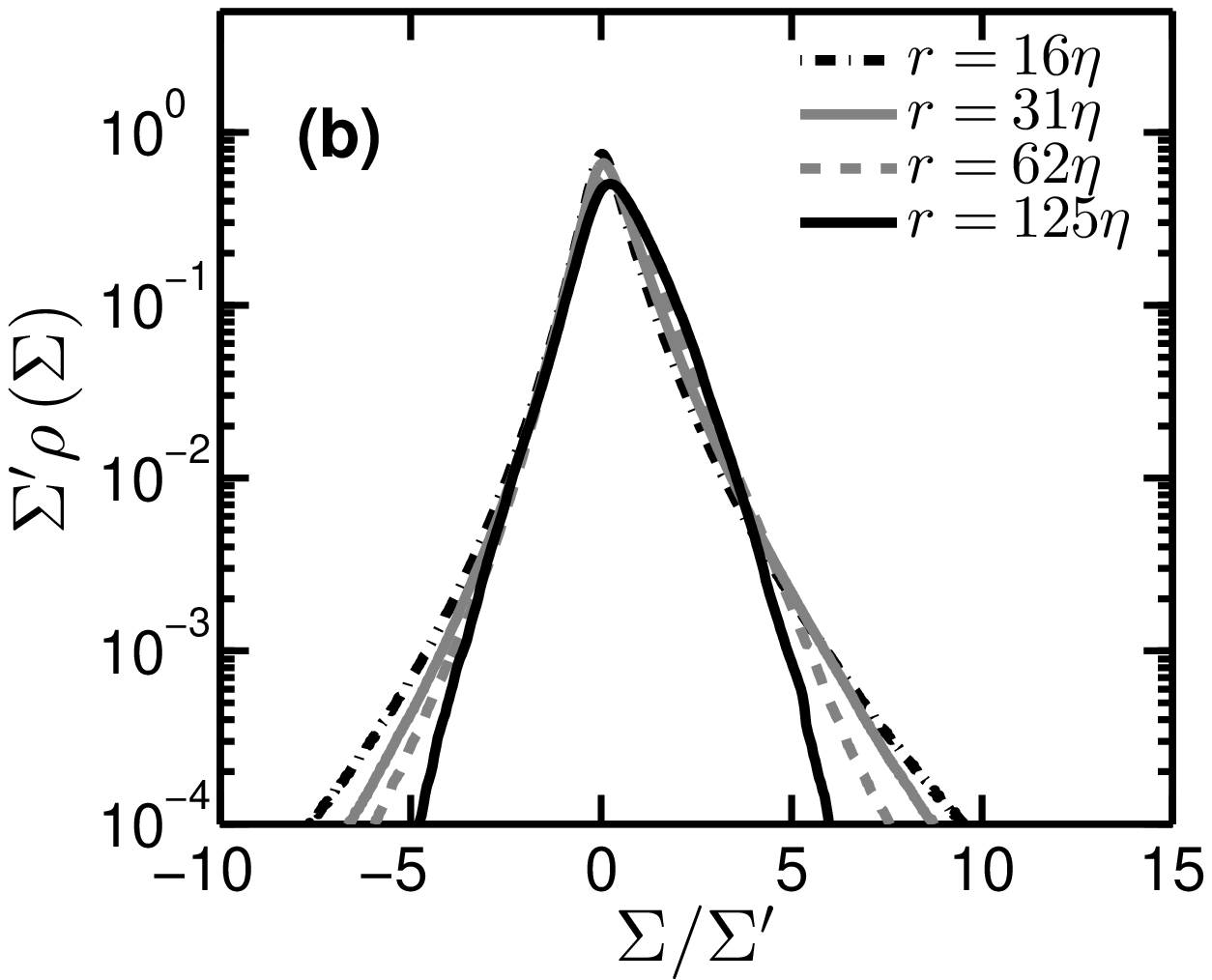}\\\includegraphics[trim = 0cm 0cm 0cm
    0cm,width=5.2cm]{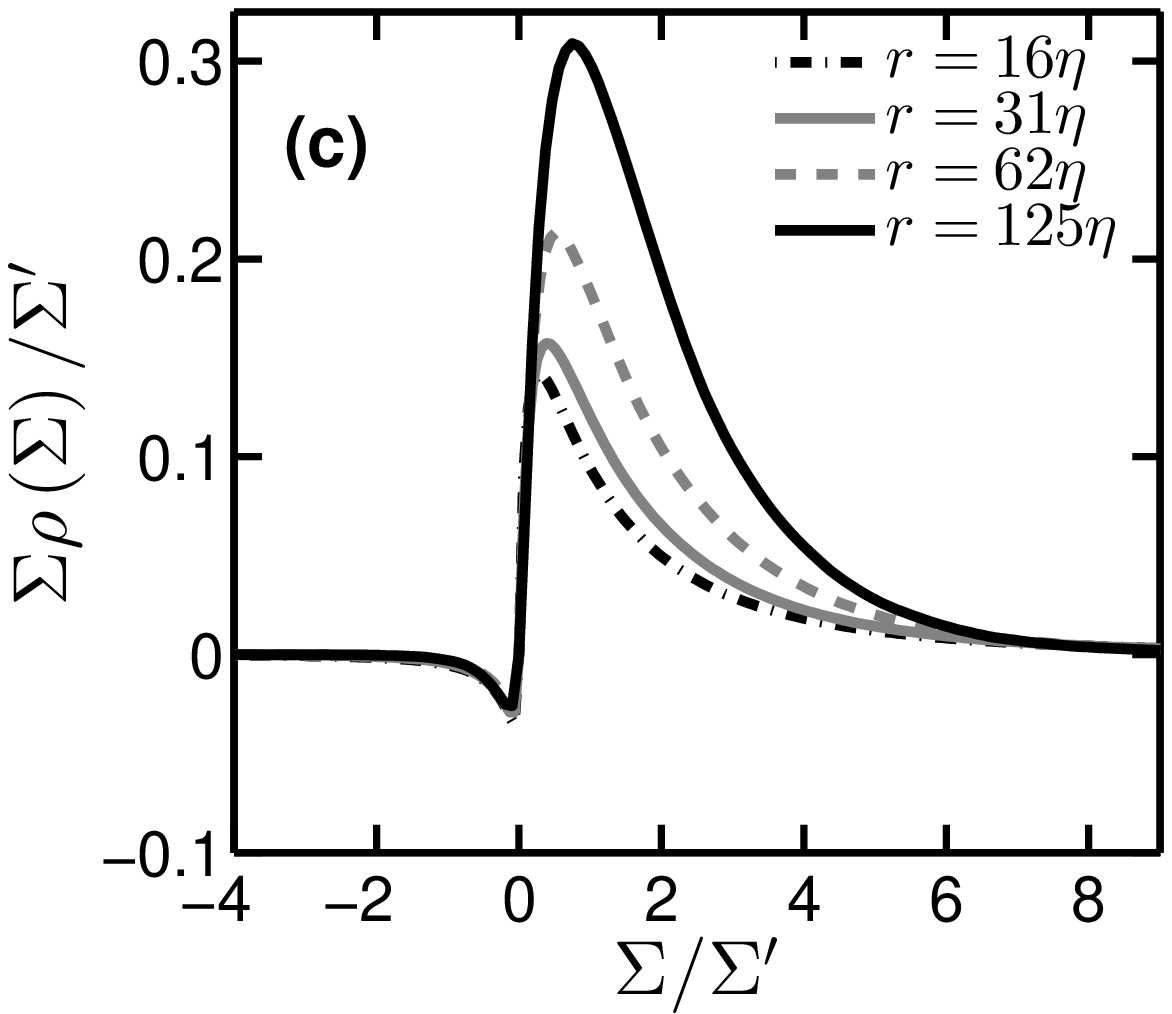}\hspace{1.5cm}\includegraphics[trim = 0cm 0cm 0cm
    0cm,width=5.2cm]{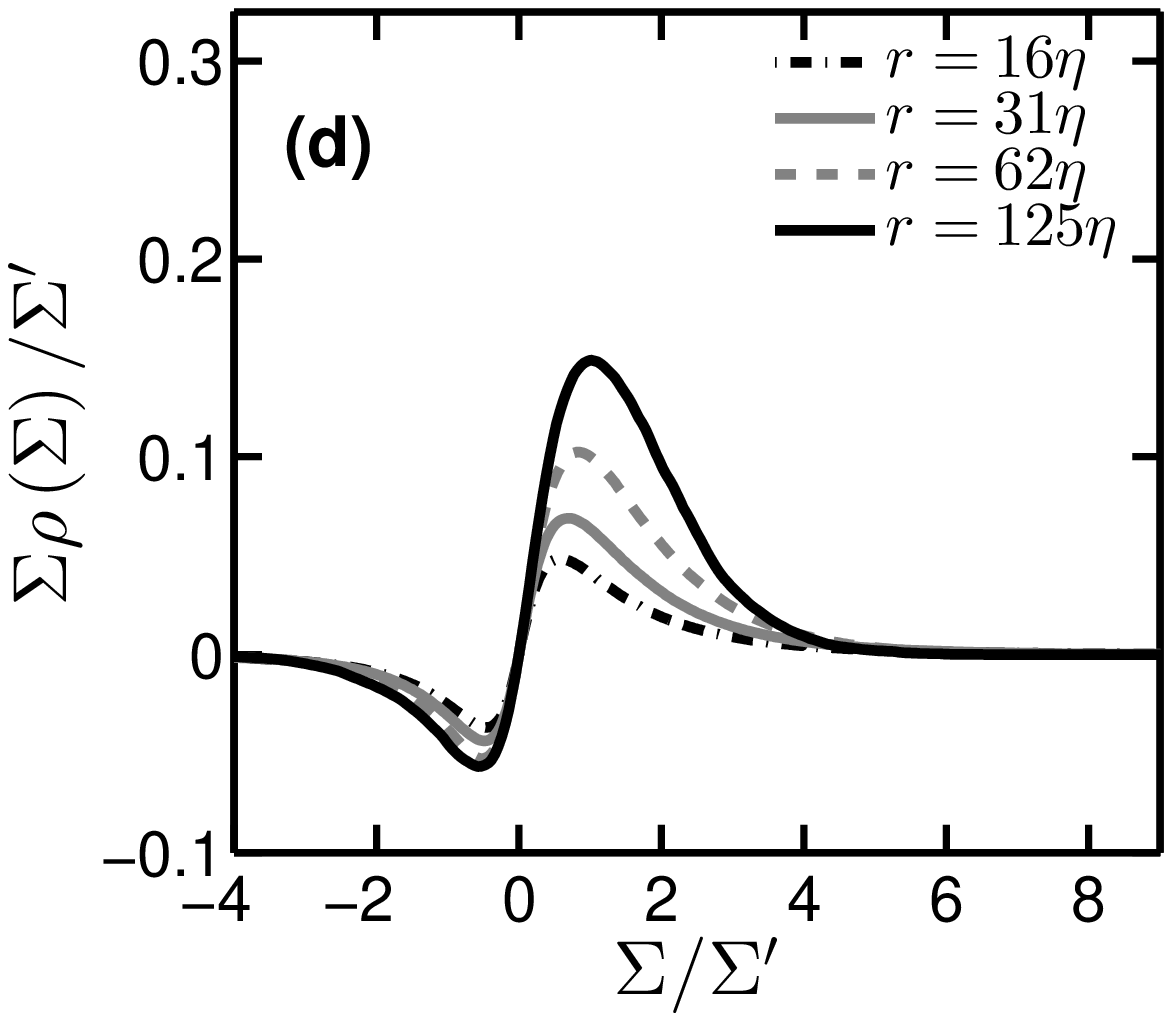}
  \end{minipage}
  \caption{Top: probability density functions (PDFs) of $\Sigma$ in HIT2, 
normalised by the standard deviation $\Sigma'$. 
    (a) Gaussian filter. (b) Sharp filter. Bottom: 
    weighted PDF, whose integral defines $\Phi_F$ and $\Phi_B$. (c) Gaussian filter. 
    (d) Sharp filter.
    \label{fig:SGSdiss_1D_PDF}}
\end{figure}

 The probability density function $\rho(\Sigma)$ can be seen 
on Fig.~\ref{fig:SGSdiss_1D_PDF}(a). Its positive skewness indicates that strong 
events are more likely when $\Sigma$ acts as a sink in Eq.~(\ref{eqn:kin_e_filt}) than 
as a source. The tails of $\rho$ become narrower with increasing filter width, in 
agreement with Ref.~\onlinecite{aoyama_2005}. 
Whereas the volume ratio of forward cascade to backscatter is known to favor 
the former over the latter for Gaussian filters,\cite{piomelli_1991,aoyama_2005} the 
ratio of the total forward energy flux 
$\Phi_{F}=\int_0^\infty\Sigma \rho\left( \Sigma \right)\dd\Sigma$ to that of backscatter 
$\Phi_{B}=\int_{-\infty}^{0}\Sigma \rho\left( \Sigma \right)\dd\Sigma$ is not documented. 
Their ratio is given on Table~\ref{tab:predom_Phi_D}, which reveals an even stronger 
preponderance of the forward cascade than what could be inferred from the 
volume ratios alone. To emphasize this point, Fig.~\ref{fig:SGSdiss_1D_PDF}(c) 
displays the value of the integrand defining $\Phi_{F}$ and $\Phi_{B}$. The sharp 
spectral filter leads to a much more symmetric picture of the cascade, as inferred 
by comparing Fig.~\ref{fig:SGSdiss_1D_PDF}(c) and (d). 
Fig.~\ref{fig:SGSdiss_1D_PDF}(b), with a sharp filter, shows that $\rho(\Sigma)$ is 
less skewed than on Fig.~\ref{fig:SGSdiss_1D_PDF}(a), with a Gaussian filter, yet 
the wider tails for decreasing $r$ occur with both filters. 
\par
\begin{figure}[t!]
  \centering
  \begin{minipage}{\textwidth}
    \includegraphics[trim = 0cm 0cm 0cm 0cm,width=5.265cm]
    {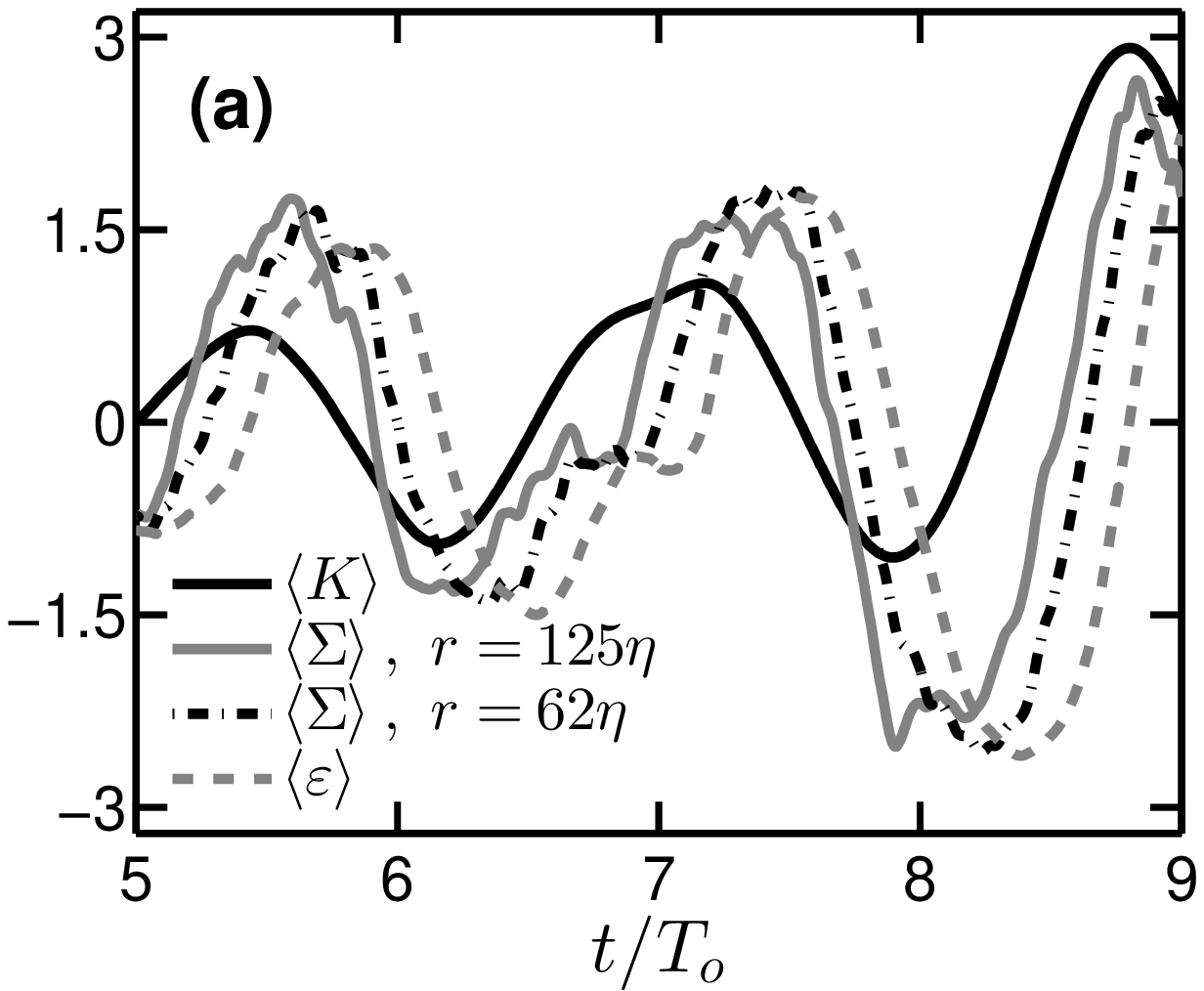}\hspace{1.5cm}\includegraphics[trim = 0cm 0cm 0cm 0cm,width=5.265cm]{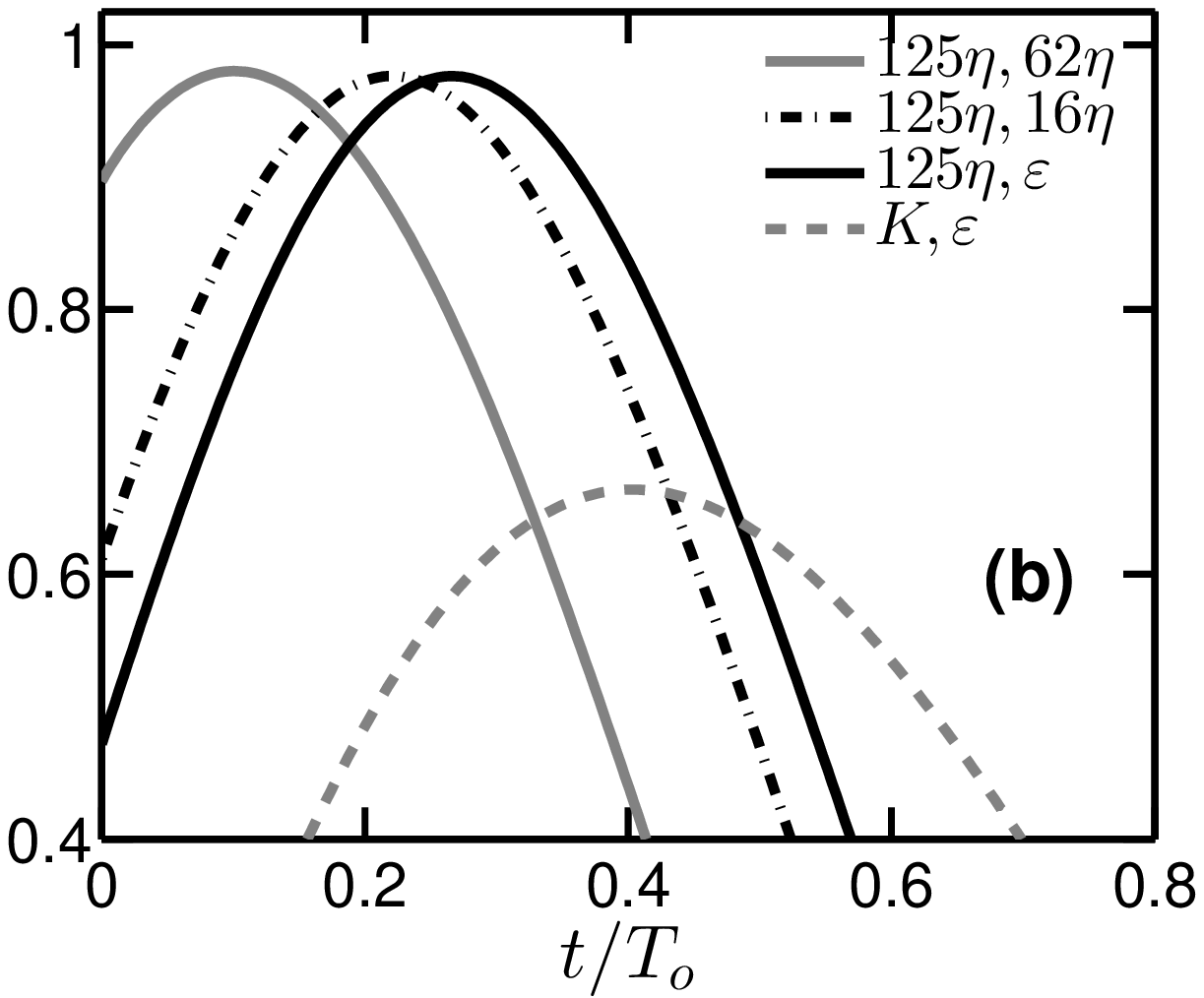}\\ \vspace{0cm}\includegraphics[trim = 0cm 0cm 0cm 0cm,width=9cm]{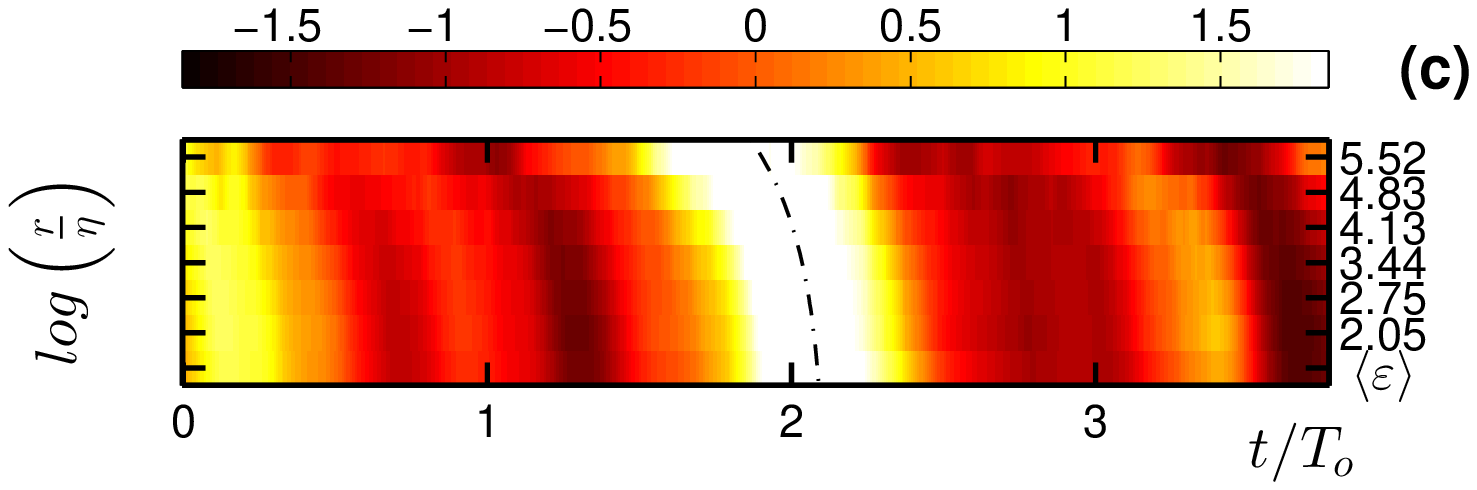}
  \end{minipage}
  \caption{(a) Temporal evolution of spatially-averaged quantities, centered and normalized by
    their standard deviation; HIT2. 
    (b) Cross-correlation curves between time series of $\mean{\Sigma}$ at
    various filter widths and between $\mean{K}$ and $\mean{\varepsilon}$; HIT2.
    (c) Time-scale diagram of $\mean{\Sigma}$, with $r$ decreasing from top 
      to bottom and $\mean{\varepsilon}$ added as the bottom band; HIT3 with $r/\eta$ 
      values from Table~\ref{tab:predom_Phi_D}. The dash-dotted line corresponds to
      $\overline{\varepsilon}^{1/3}\Delta t = (250\eta)^{2/3}-r^{2/3}$ - see Eq.~(\ref{eqn:2_3rds}).
    \label{fig:S_time_evo}}
\end{figure}

 We now move towards the dynamics of $\mean{\Sigma}$, based on its
time series which we computed for all flows with a
Gaussian filter - the sharp filter was used only on a few HIT fields widely spaced in time.
They are shown on Fig.~\ref{fig:S_time_evo}(a) at two filter 
widths, together with the kinetic energy $\mean{K}=\mean{u_iu_i}/2$ and the viscous 
dissipation $\mean{\varepsilon}$. It is clear that the signals are correlated, but that 
there is a delay separating them. $\mean{K}$ and $\mean{\varepsilon}$ behave as the 
earliest and latest signals, while the delay of $\mean{\Sigma}$ with respect to $\mean{K}$
increases with decreasing $r$. 
To visualize this effect more clearly,
Fig.~\ref{fig:S_time_evo}(c) displays the temporal evolution of $\mean{\Sigma}$ as a
color map where the abscissae are time. Color bands corresponding to $\mean{\Sigma}$
have been ordered vertically with $r$ decreasing logarithmically downwards, and
$\mean{\varepsilon}$ added at the bottom. The propagation across $r$ and $t$ of
disturbances in $\mean{\Sigma}$ is evident. To quantify this process, we compute the
temporal cross-correlation of all these signals with each other, as well as their
temporal autocorrelation. A few such correlations are illustrated on 
Fig.~\ref{fig:S_time_evo}(b), where the peak appears at the average delay
between the two chosen signals. We start by looking at $\Delta t_{tot}$, the average 
time taken by a change in $\mean{K}$ to propagate and appear as a change in 
$\mean{\varepsilon}$. It is compiled in Table~\ref{tab:DNS_settings} for all our flows, 
which shows that $\Delta t_{tot}$ is approximately half the integral dissipation time 
$T_{o}\equiv\overline{K}/\overline{\varepsilon}$. Similar data are scarce in the 
literature, so that a comparison is not straightforward. In a computational study 
of homogeneous shear flow,\cite{PumirHST} the delay between the time histories of 
$\mean{K}$ and $\mean{\varepsilon}$ was estimated to be of the order of 
$\Delta t_{tot}/T_{o}\approx 1.5$ at $Re_{\lambda}\approx 50$. Since this disagrees 
with our $\Delta t_{tot}/T_{o}=0.49$ at $Re_{\lambda}\approx 107$, we repeated the 
simulation in Ref.~\onlinecite{PumirHST}, and found $\Delta t_{tot}/T_{o}= 0.52$. 
The discrepancy can probably be attributed to the different estimation methods. 
Cross-correlations were not computed in Ref.~\onlinecite{PumirHST}, where 
$\Delta t_{tot}$ was not the main focus of the study. \textcolor{black}{A value of 
$\Delta t_{tot}/T_{o}= 0.44$ can be extracted from the data in 
Ref.~\onlinecite{Horiuti_PoF2013} where a DNS of HIT was ran at $Re_{\lambda}=122$, which 
agrees well with our findings.}
In Ref.~\onlinecite{PearsonPRE}, 
a $\Delta t_{tot}$ was defined as the delay between $\mean{K}^{3/2}/L_{int}$ and $\mean{\varepsilon}$, where
$L_{int}=(3\pi/4\mean{K})\int k^{-1}E(k)\dd k$. Using temporal cross-correlations in
HIT at $Re_{\lambda}=219$ -- comparable to our HIT2, they found 
$\Delta t_{tot}/T_{o}\approx 0.21$. We found $\Delta t_{tot}/T_{o}\approx 0.28$
using our HIT2 and the same definition of $\Delta t_{tot}$ as in 
Ref.~\onlinecite{PearsonPRE}, confirming that the lag between $\mean{K}^{3/2}/L_{int}$ and 
$\mean{\varepsilon}$ is shorter than that between $\mean{K}$ and $\mean{\varepsilon}$. 
\par A comment on the origin of the time dependence of $\mean{K}$, $\mean{\varepsilon}$ 
and $\mean{\Sigma}$ is in order. The temporal oscillations of $\mean{K}$ and 
$\mean{\varepsilon}$ in HST are known to be physically caused and related to bursting 
.\cite{PumirHST} In the 
HIT simulations, the turbulence is sustained by a deterministic force
\begin{equation}
  \hat{f}_{i}\left(\boldsymbol{k},t\right)=
  \begin{cases} 
    \overline{\varepsilon} \hat{u}_{i}\left(\boldsymbol{k},t\right)/\left[2E_{f}(t)\right], 
    & \text{if $0 < k < k_{f}$},\\
    0, & \text{otherwise},
  \end{cases}
\end{equation}
where $\overline{\varepsilon}$ is the target mean dissipation,\cite{Machiels_PRL1997}
$E_{f}(t) = \int_{0}^{k_{f}}E\left(k,t
\right) \dd k$ and $k_{f}=4\pi/L_{x}$. This commonly used scheme is 
mildly unstable, because of the delay between $\hat{f}_{i}$ and $\mean{\varepsilon}$. It generates 
time oscillations of the energy while maintaining a constant resolution of 
$k_{max}\eta=1.5$ in the mean. 
\begin{table}[b!]
\vspace{-0.7cm}
\caption{Symbol legend for Figs.~\ref{fig:dt_all_vs_r} and \ref{fig:2_3rds}. 
$r_{a}/\eta=10a\sqrt{6}/\pi$ so that $r_{_{1}}\approx8\eta$, $r_{_{2}}\approx16\eta$, 
\textit{etc.}  \label{tab:legend_figs}}
\begin{ruledtabular}
\begin{tabular}{lcccccc}
  & $r \rightarrow \mean{\varepsilon}$ & 
  $a = 1$ & 
  $a = 2$ & 
  $a = 4$ & 
  $a = 8$ & 
  $a = 16$ \\ \hline
HST & + & $N/A$ & $\times$ & $\ast$ & $\Diamond$ & $N/A$ \\
HIT1 & $\square$ & $\vartriangle$ & $\Circle$ & $\triangledown$ & $N/A$ & $N/A$ \\
HIT2 & $\color{gray}\blacksquare$ & $\color{gray}\blacktriangle$ & $\color{gray}\CIRCLE$ 
& $\color{gray}\blacktriangledown$ & $\color{gray}\blacktriangleright$ & $N/A$ \\
HIT3 & $\blacksquare$ & $\blacktriangle$ & $\CIRCLE$ & $\blacktriangledown$ 
& $\blacktriangleright$ & $\blacktriangleleft$ \\ 
\end{tabular}
\end{ruledtabular}
\end{table}
With these two flows at hand, we can assess the dependence 
of $\Delta t_{tot}$ on the large-scale forcing. We define the characteristic time scale 
$T_{KK}$ of the kinetic energy as the width of the temporal autocorrelation of $\mean{K}$ 
at half its peak height. The ratio $T_{KK}/\Delta t_{tot}$ is between 2 and 2.5 for all 
our HIT simulations, but about 9 for the HST -- see Table~\ref{tab:DNS_settings}. Yet 
changes in $\mean{K}$ 
appear as changes in $\mean{\varepsilon}$ within half a large-eddy turnover time in the 
two differently forced flows, suggesting that $\Delta t_{tot}/T_{o}$ is \textcolor{black}{
a common feature 
of the energy cascade when the large scales fluctuate with periods in our range of 
$T_{KK}/T_{o}$. The dependence of our measured $\Delta t_{tot}/T_{o}$ on the large-scale period
could be studied further by extending this range of $T_{KK}/T_{o}$ with the addition of a 
modulating frequency in the forcing, as done Ref.~\onlinecite{kuczaj2006}, or with 
a stochastic forcing.\cite{eswaran1988}} 
\par
\begin{figure}[t!]
  \centering
  \begin{minipage}{\textwidth}
    \includegraphics[trim = 0cm 0cm 0cm 0cm,height=4.58cm]{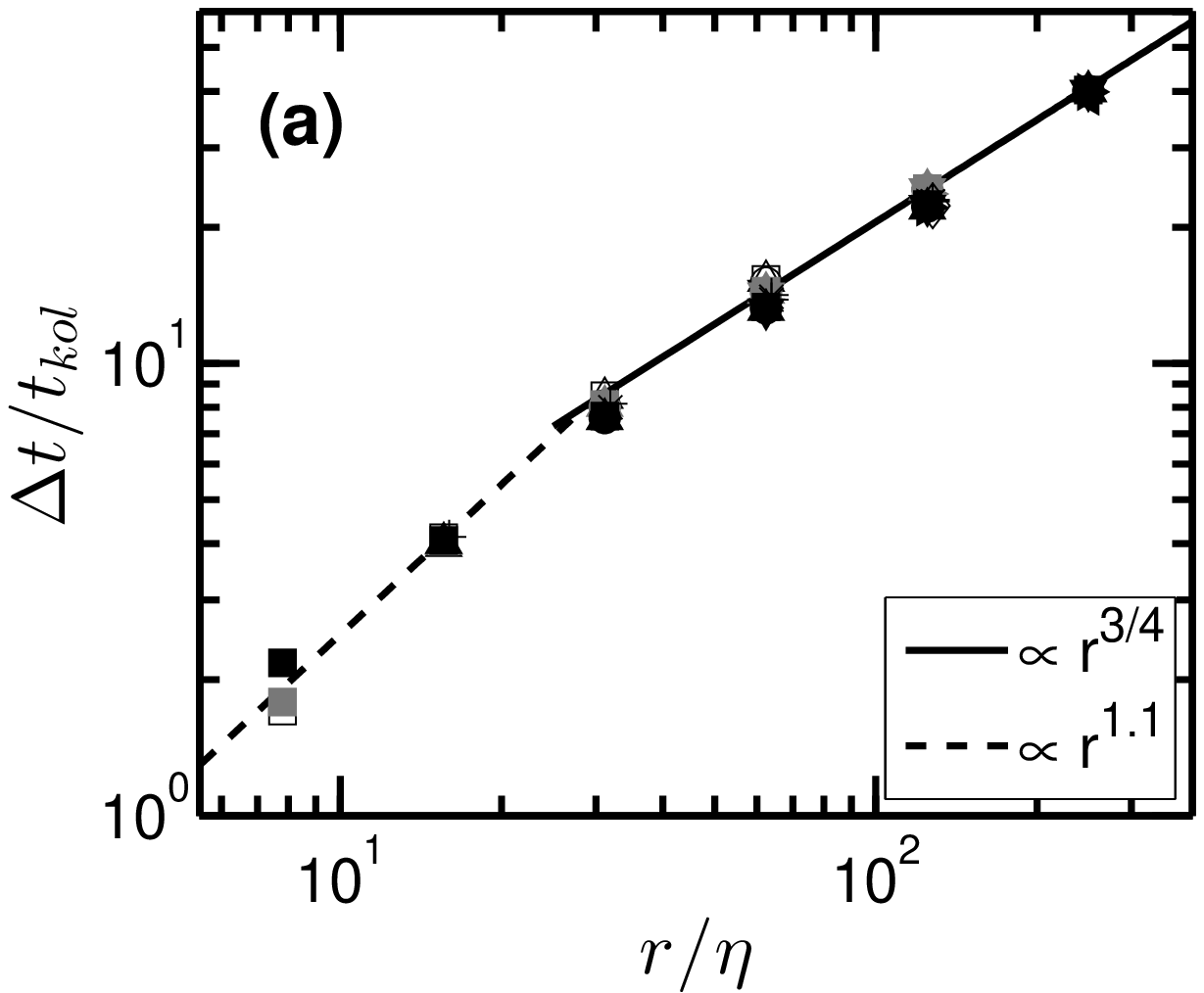}\hspace{1.5cm}\includegraphics[trim = 0cm 0cm 0cm 0cm,height=4.455cm]{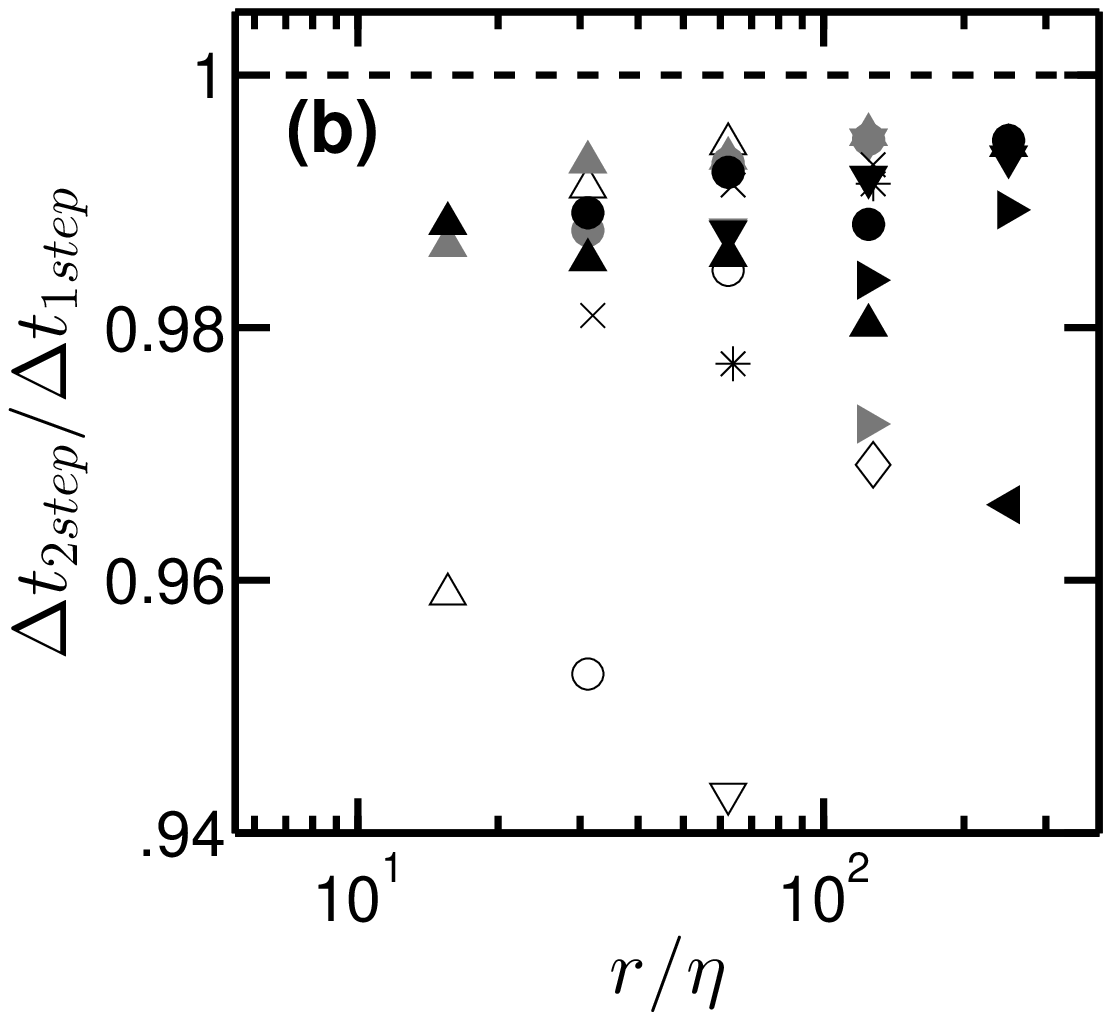}
    \end{minipage}
  \caption{(a) Delay between $\mean{\Sigma}$ and $\mean{\varepsilon}$, measured either as 
the one-step delay, or as the sum of two intermediate steps involving $r_{a}$. 
(b) Ratio between two- and one-step delays. See Table~\ref{tab:legend_figs} for symbols 
of $r_a$ and flow.  
  }\label{fig:dt_all_vs_r}
\end{figure}

 We next test the additivity of the delay times. We want to see if the time taken by a
disturbance in the energy flux in going from scale size $r_{3}$ to $r_{1}$ is equal
to the sum of the delays in going from $r_{3}$ to $r_{2}$ and from $r_{2}$ to $r_{1}$,
where $r_{3} > r_{2} > r_{1}$. Fig.~\ref{fig:dt_all_vs_r}(a) displays the average
time needed for disturbances in the energy fluxes at a given scale $r$ to travel down
to $\mean{\varepsilon}$. This time is computed for all available combinations of two
intermediate jumps starting at $r$ and ending in $\mean{\varepsilon}$. The agreement
between the different jump combinations within the same flow and across
different flows is satisfactory. In order to highlight any residual discrepancy, we
display the value of the ratios between one- and two-step cascading times on
Fig.~\ref{fig:dt_all_vs_r}(b). A ratio close to unity implies additivity of the
delays, which is confirmed for all the starting scales $r$ and flows examined. 
Note the narrow range of the vertical axis. Note also that 
Fig.~\ref{fig:dt_all_vs_r}(a) hints at two different regimes for $r$ above and below 
approximately $30\eta$. This is consistent with the results of 
Ref.~\onlinecite{yoshida2005regeneration}, who showed that viscous eddies below 
$r/\eta \approx 30$ are enslaved to larger ones above that scale. In essence, 
$r/\eta = 30$ is the lower limit of the inertial cascade.
\par In an influential paper, Lumley discusses two
cascade models,\cite{Lumley_somethoughts_PoF1992} 
and proposes a forcing experiment similar to the present one to
distinguish between them. He starts by considering a hierarchy of discrete eddies of
decreasing size. In the first model, each eddy transfers its energy to the one
immediately below. The transfer occurs at a rate determined by the corresponding
scale-dependent eddy turnover time, $[k^3 E(k)]^{-1/2}\sim k^{-2/3}\sim r^{2/3}$, so that the 
propagation of energy from the large towards the small scales develops into a 
front-like diffusion through scale-space with a finite scale-dependent velocity. 
In the second model, most of the energy is still transferred to the immediately 
smaller eddy below, but a fraction is passed to other eddies further along the cascade. 
Hence the smallest eddies receive a small amount of energy almost immediately after 
it is injected into the system, and increasing amounts as time goes on. The 
difference between the two models is that all the energy in the first one has to 
pass through each eddy size, resulting in additive cascade times, while this 
additivity is not guaranteed in the second model. Theoretical arguments were put forward 
both for and against long-range energy transfer,\cite{YeungBrass91,doma93} 
and attempts were made to carry out the forcing experiment, but they were hindered by 
the low Reynolds numbers available from simulations at the time. The matter has 
remained controversial until now, and our additive data on Fig.~\ref{fig:dt_all_vs_r}
favors the local model. 
\par
\begin{figure}[t!]
  \centering
  \begin{minipage}{\textwidth}
    \includegraphics[trim = 0cm 0cm 0cm 0cm,height=4.455cm]{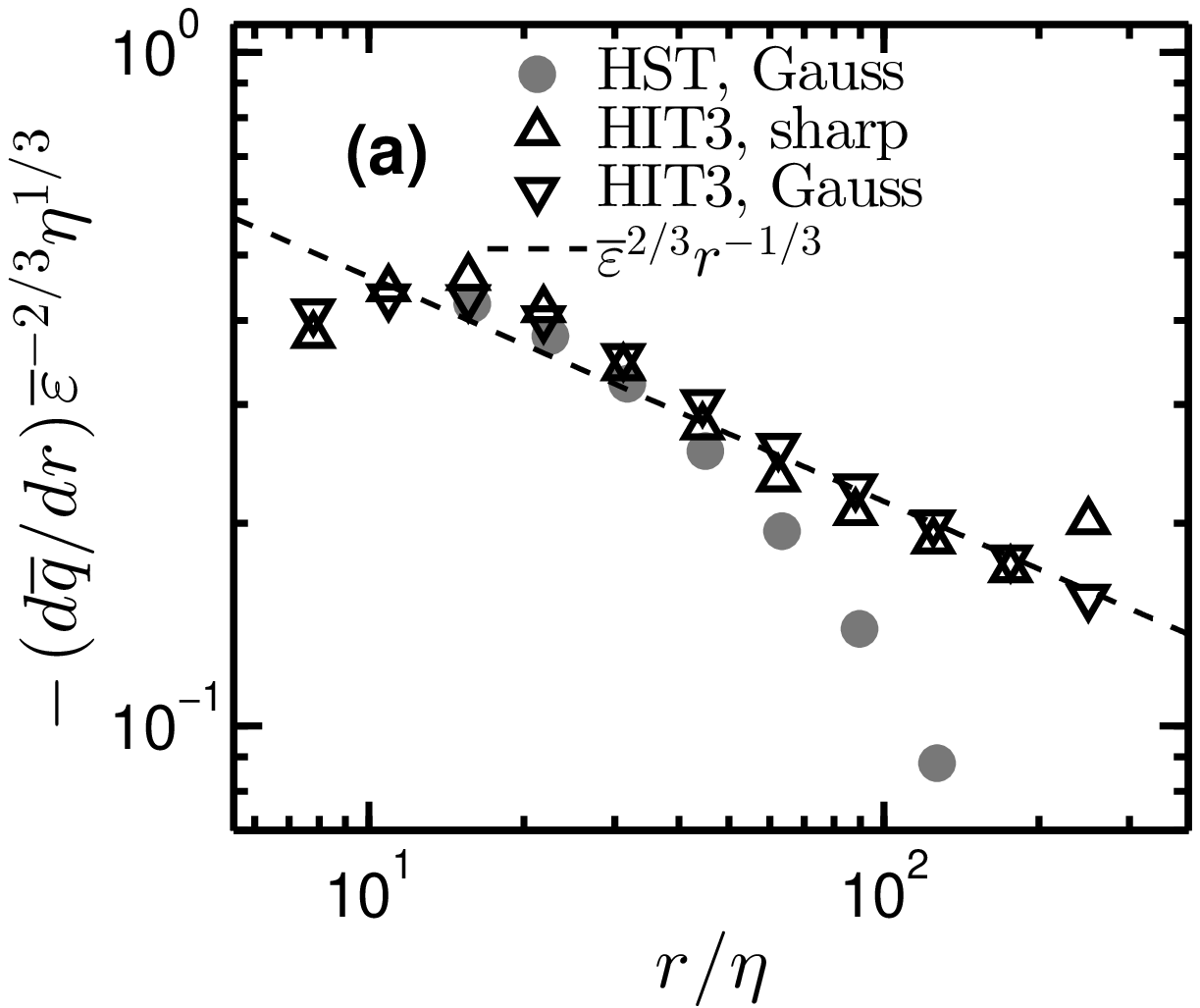}\hspace{1.5cm}\includegraphics[trim = 0cm 0cm 0cm 0cm,height=4.455cm]{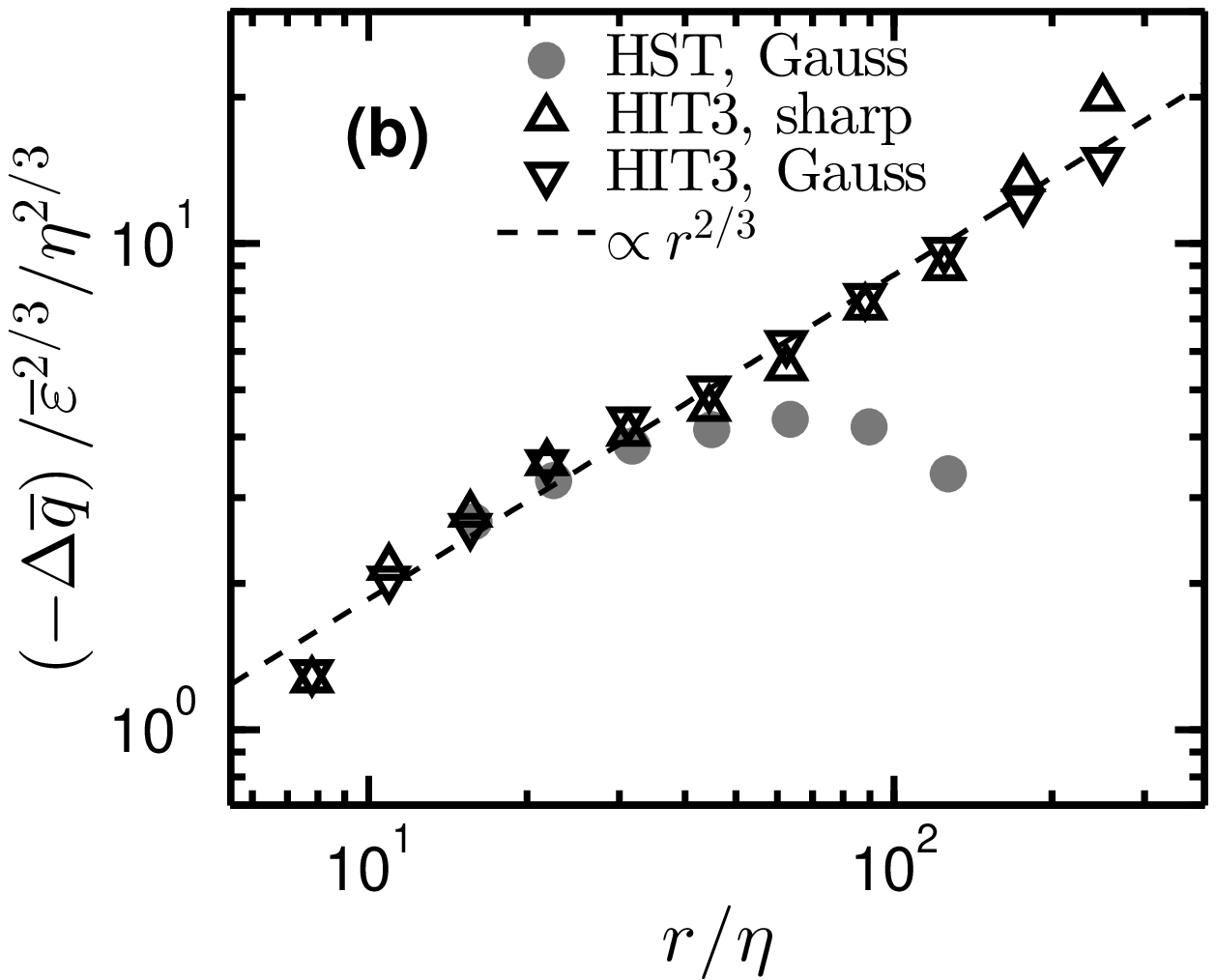}
  \end{minipage}
\caption{(a) Derivative of 
$\overline{q}=\overline{\widetilde{u}_{i}\widetilde{u}_{i}}/2$ with respect to $r$. The dashed line corresponds to
$\overline{\varepsilon}^{2/3}r^{-1/3}$.
(b) Energy content within a band of scales between $r$ and $r+\Delta r$, where 
$\Delta r$ goes from a given
$r$ to the next bigger $r$ in the plotted series. 
$\Delta \overline{q}=\frac{1}{2}\overline{\widetilde{u}_{i}\widetilde{u}_{i}}(r+\Delta r)-
\frac{1}{2}\overline{\widetilde{u}_{i}\widetilde{u}_{i}}(r)$.}\label{fig:assumptions}
\end{figure}

 We now focus on the scaling of $\Delta t$.
We start by introducing the strong assumption that
the values of $\Delta t$ we measure are 
between scales $r$ within an inertial range, where
$r$ and $\overline{\varepsilon}$ are the only
relevant quantities.
Within this simplistic framework, then, a velocity in $r$-space 
can be defined as
\begin{equation}
  \dot{r} = \overline{\varepsilon}/\overline{\rho}_{_{E}}, \label{eqn:ODE}
\end{equation}
where the energy density $\overline{\rho}_{_{E}}(r)$ 
is a real-space equivalent of $\overline{E}(k)$. We put forward the 
following candidate for $\overline{\rho}_{_{E}}$, based on 
$q = \widetilde{u}_{i}\widetilde{u}_{i}/2$:
\begin{equation}
  \overline{\rho}_{_{E}} = -\frac{d\overline{q}}{dr}.
\end{equation}
Other densities have been introduced in physical space.
Townsend's $r$-derivative of the correlation function, 
\cite{townsend1980structure} or the signature function found in 
Ref.~\onlinecite{davidson2005identifying} are two examples.
We chose our expression as it is based on the filtering approach
we use. The data on  
Fig.~\ref{fig:assumptions}(a) shows that the energy density 
$-d\overline{q}/dr$ 
in our two flows is a positive quantity. 
In Fig.~\ref{fig:assumptions}(b) 
we see that the energy 
$-\Delta \overline{q}$ 
contained between $r$ and $r+\Delta r$ is a quantity which grows 
proportionaly to 
$r^{2/3}$ within a reasonable range in HIT3 - not so in HST with
much smaller scale separation.
Such $r^{2/3}$ behaviour is consistent
with the Kolmogorov-Obukhov theory,
\cite{K41,obukhoff1941energy}
yet it is based on a completely different
flow decomposition from the structure function.
\begin{figure}[h!]
  \includegraphics[trim = 0cm 0cm 0cm 0cm,height=4.455cm]{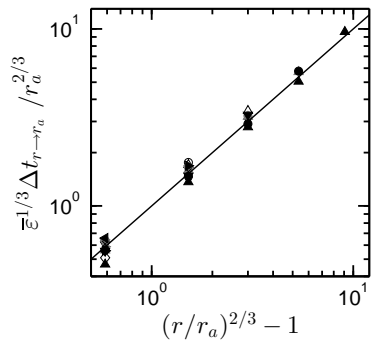}
\caption{$\overline{\varepsilon}^{1/3}\Delta t_{r\rightarrow r_{a}}/r_{a}^{2/3}$
against $(r/r_{a})^{2/3}-1$, where $\Delta t_{r\rightarrow r_{a}}$ is the 
average delay
between $\mean{\Sigma(r)}$ and $\mean{\Sigma(r_{a})}$, with $r>r_{a}$. Symbols as in
Table~\ref{tab:legend_figs}. The solid line corresponds to 
Eq.~(\ref{eqn:2_3rds})}\label{fig:2_3rds}
\end{figure}
Substituting $\overline{\rho}_{_{E}}$ in Eq.~(\ref{eqn:ODE}) by 
$\overline{\varepsilon}^{2/3}r^{-1/3}$, and integrating from $r_{a}$ to $r$
leads to 
\begin{equation}
  \Delta t_{r\rightarrow r_{a}}=\overline{\varepsilon}^{-1/3}\left(r^{2/3}-r^{2/3}_{a}\right),
  \label{eqn:2_3rds}
\end{equation}
which implies that our data should fall on a straight line when plotted 
logarithmically as done in Fig.~\ref{fig:2_3rds}. The agreement 
is not completely unsatisfactory. Particularly 
when considering
the inertial range assumptions used which 
have no reason to apply if $r$ or $r_{a}$ are
below the viscous limit of $30\eta$, 
or given the poor compliance of HST to the inertial range
scaling with $r$ - see Fig.~\ref{fig:assumptions}. A dashed line following 
Eq.~(\ref{eqn:2_3rds}) for HIT3 was added on Fig.~\ref{fig:S_time_evo}(c).
\par A derivation of Eq.~(\ref{eqn:2_3rds}) carried out
in spectral space can be found in Ref.~\onlinecite{Pope_book}, leading to a
$k^{-2/3}$ dependence of $\Delta t$. Ref.~\onlinecite{Wan_Meneveau} 
studied Lagrangian time correlations of
$\Sigma(k)$ and $\varepsilon$, which hinted at a $k^{-2/3}$ 
dependence of the peaks in their correlations - see inset of their Fig.~2. 
Earlier, the same group measured the 
temporal correlation between the energy at a given scale and the energy at a 
smaller scale found later by following the flow in both forced and decaying HIT.
\cite{MeneveauLund} 
Their conclusion that the peak in correlation happens later for increasing scale 
separation is consistent with what we observe. A difficulty with that work was 
the use of correlations of energy rather than energy flux. Fluxes are the quantities
conserved across cascades, and the natural objects for their study.
Furthermore, they could only consider one-jump delays, ruling out the additivity test
on Fig.~\ref{fig:dt_all_vs_r} which supports the locality of the energy cascade
in an average sense. 
\vspace{-0.4cm}
\begin{acknowledgments}
\vspace{-0.3cm}
This work was supported by the Multiflow grant
ERC-2010.AdG-20100224. Computational time was provided 
on GPU clusters at the BSC (Spain)
under projects FI-2014-2-0011, FI-2015-1-0001 and in Tianjin's
NSC (China).
\end{acknowledgments}
\bibliography{references}
\end{document}